\documentstyle[12pt]{article}
\newcommand{\bfr}{\begin{flushright}}
\newcommand{\efr}{\end{flushright}}
\newcommand{\bfl}{\begin{flushleft}}
\newcommand{\efl}{\end{flushleft}}
\begin{document}
\begin{center}
\thispagestyle{empty}
\null
\vskip-1truecm
hep-th/9505037
\rightline{IC/95/61}
\rightline{OTSUMA--HEP--9502}
\vskip1truecm
International Atomic Energy Agency\\
and\\
United Nations Educational Scientific and Cultural Organization\\
\medskip
INTERNATIONAL CENTRE FOR THEORETICAL PHYSICS\\
\vskip1.5truecm
{\bf THE ELECTRICALLY CHARGED BTZ BLACK HOLE\\
WITH SELF (ANTI--SELF) DUAL MAXWELL FIELD}\\
\vspace{1.5cm}
Masaru Kamata\footnote
{\normalsize On leave of absence from: Kisarazu National College of Technology,
2-11-1 Kiyomidai-Higashi, Kisarazu, Chiba 292, Japan.\\
E-mail addresses: kamata@ictp.trieste.it; kamata@gokumi.j.kisarazu.ac.jp
(after July
1995)}\\
International Centre for Theoretical Physics, Trieste, Italy\\
\bigskip
and\\
\bigskip
Takao Koikawa\footnote
{\normalsize E-mail address: koikawa@csc.otsuma.ac.jp}\\
School of Social Information Studies, Otsuma Women's University,\\
Karakida, Tama, Tokyo 206, Japan.\\
\end{center}
\vspace{0.5cm}
\centerline{ABSTRACT}
\bigskip

The Einstein-Maxwell equations with a negative cosmological constant
$ \Lambda $ in $ 2+1 $ spacetime dimensions discussed by Ba\~{n}ados,
Teitelboim and Zanelli are solved by assuming a
self (anti-self) dual equation $ E_{\hat r} = \pm B_{\hat{~}} $, which is
imposed on the orthonormal basis components of the electric field $ E_{\hat r}
$
and the magnetic field $ B_{\hat{~}} $.
This solution describes an electrically charged extreme black hole with mass
$ M = 8{\pi}GQ_{e}^2 $, angular momentum $ J = \pm
8{\pi}GQ_{e}^2/|\Lambda|^{1/2} $ and  electric charge $ Q_{e} $. Although the
coordinate components of the electric field $ E_{r}
$ and the magnetic field $ B $ have singularities on the horizon at $ r =
(4{\pi}GQ_{e}^2/|\Lambda|)^{1/2} $, the spacetime has the same
value of constant negative curvature
$ R = 6\Lambda $ as that of Ba\~{n}ados et al.
\vspace{0.7cm}
\begin{center}
MIRAMARE --TRIESTE\\
\medskip
April 1995
\end{center}


\newpage

Lower dimensional gravity theories have been attracting considerable attention
in the hope that they provide some clues for solving some of the problems in
the
quantum  gravity theories in the four-dimensional spacetime.

Recently Ba\~{n}ados, Teitelboim and Zanelli (BTZ) \cite{BTZ,BHTZ} found that
the $2+1$ dimensional Einstein theory with a negative cosmological constant
$\Lambda$ admits a  solution which has almost all of the features of the black
hole with mass $M$ and angular  momentum $J$.
Although this solution does not approach any flat spacetime asymptotically
due to
its constant negative  curvature $ R=6\Lambda $, it has an
event horizon, an ergosphere when $J\neq0$, and nonzero Hawking temperature.

In this Letter we present an exact electrically charged BTZ black hole solution
to the Einstein-Maxwell equations in $2+1$ spacetime dimensions,\footnote[1]
{The authors of \cite{BTZ,BHTZ} also mentioned in their papers an electrically
charged rotating solution, in which $ A=-Q_e\ln(r/r_0)dt $. However their
solution with nonzero  angular momentum $J$ and nonzero electric charge $Q_{e}$
is not a solution because  the azimuthal $\phi $ component of the Maxwell
equation (\ref{eq:Max}) is not zero but $  JQ_{e}/r^3 $.}
which is obtained by assuming a self dual (SD) (anti-self dual (ASD)) equation
\footnote[2]
{In fact by suppressing the index $ {\hat r} $ we can rewrite (\ref{eq:SD})
into the
self dual (anti-self dual) equation in the $ (t, \phi) $ space: $
{}^*f_{\hat\mu} = \varepsilon
f_{\hat\mu}, $ where $ {}^* f_{\hat\mu} \equiv
\varepsilon_{{\hat\mu}{\hat\nu}} f^{\hat\nu} $
is the dual of $ f_{\hat\mu} \equiv F_{{\hat\mu}{\hat r}}$ with $
\varepsilon_{{\hat\mu}{\hat\nu}} $ the completely antisymmetric symbol
normalized as $
\varepsilon^{{\hat t}{\hat \phi}} \equiv 1 $ and the indices run over only
$ {\hat t} $ and $ {\hat\phi} $.}
\begin{equation}
E_{\hat r} = \varepsilon B_{\hat{~}},\hspace{5mm} \varepsilon = \pm 1,
\label{eq:SD}
\end{equation}
where $ E_{\hat r} \equiv F_{\hat{t}\hat{r}} $ and $ B_{\hat{~}} \equiv
F_{\hat{r}\hat{\phi}} $
are
the orthonormal basis components of the electric field and the magnetic field,
respectively.
The electric field $ E_{\hat r} $ and the magnetic field $ B_{\hat{~}} $
which is a
pseudoscalar in the $2+1$ dimensional spacetime form a ``Poynting
pseudovector'' $
E_{\hat r}B_{\hat{~}} $, and this gives the black hole a nonzero angular
momentum $ J $. We will show in the following that the Einstein-Maxwell
second-order
differential equations are reduced to the first-order ones under the SD
(ASD) equation
(\ref{eq:SD}) and these reduced differential equations can be solved by simple
quadratures.

The Einstein-Maxwell action is
\begin{equation}
S=\frac{1}{16{\pi}G}\int
\sqrt{-g}(R-2\Lambda-4{\pi}GF^2)d^3x,
\label{eq:EMaction}
\end{equation}
where $G$ is Newton's constant, $\Lambda$ is the negative cosmological
constant and
$F^2 \equiv
\linebreak
g^{\mu\nu}g^{\rho\sigma}F_{\mu\rho}F_{\nu\sigma}$. The Einstein equation is
given by
\begin{equation}
G_{\mu\nu}+{\Lambda}g_{\mu\nu}=8\pi GT_{\mu\nu}, \label{eq:Ein}
\end{equation}
where $ T_{\mu\nu} $ is the energy-momentum tensor of the electromagnetic
field:
\begin{equation}
T_{\mu\nu}
=F_{\mu\rho}F_{\nu\sigma}g^{\rho\sigma}
-\frac{1}{4}g_{\mu\nu}F^2,
\label{eq:T..}
\end{equation}
whereas the Maxwell equation is
\begin{equation}
\partial_{\rho}(\sqrt{-g}g^{\mu\nu}g^{\rho\sigma}F_{\nu\sigma})=0.
\label{eq:Max}
\end{equation}

We here adopt the following line element compatible with the stationary
axisymmetric
spacetime \cite{Chandra}:
\begin{equation}
ds^2=-N^2dt^2+L^{-2}dr^2+K^2(N^{\phi}dt+d\phi)^2, \label{eq:lineelement}
\end{equation}
where $ N $, $ L $, $ K $ and $ N^\phi $ are functions of only r. Note that
any coordinate
condition such as $ K= r $ is not imposed on this. If we
set $ L = N $ and $ K = r $, the line element (\ref{eq:lineelement}) agrees
with that of BTZ
\cite{BTZ}, in which the perimeter length of the
circle with $ t = $ const. and $ r = $ const. is given by the usual formula
$2{\pi}r$.
These restrictions are, however, too stringent to derive our electrically
charged rotating
solution.
If we introduce
an orthonormal 1-form basis
\begin{equation}
\omega^{\hat{t}} \equiv Ndt, \hspace{5mm} \omega^{\hat{r}} \equiv L^{-1}dr,
\hspace{5mm} \omega^{\hat{\phi}} \equiv K(N^{\phi}dt+d\phi) \label{eq:obasis}
\end{equation}
and
form a linear combination
$ G^{\hat{t}}_{~\hat{t}}-G^{\hat{r}}_{~\hat{r}} =
8{\pi}G(T^{\hat{t}}_{~\hat{t}}-
T^{\hat{r}}_{~\hat{r}}) $ from the Einstein equation,
we have the following equation
\begin{equation}
K^{-1}K'(N^2{L^2}'-L^2{N^2}')+2[(K^{-1}K')'+(K^{-1}K')^2]N^2L^2 =-
16{\pi}GN^2B_{\hat{~}}^2.
\label{eq:Gtt-Grr}
\end{equation}
The left-hand side of this equation vanishes when $ L = N $ and $ K = r $
and so we
have $ B_{\hat{~}}=0 $, which implies $ E_{\hat r} = 0 $ because of the SD
(ASD)
equation (\ref{eq:SD}).

Since we have two Killing vectors $ \partial/\partial t $ and $
\partial/\partial \phi $ in the
stationary axisymmetric spacetime, the Maxwell equation (\ref{eq:Max}) is
immediately
solved as follows:
\begin{equation}
E_{\hat r} \equiv F_{\hat{t}\hat{r}} = \frac{C_1}{K}, \hspace{1cm}
B_{\hat{~}} \equiv
F_{\hat{r}\hat{\phi}} = \frac{C_1N^{\phi}+C_2}{N}, \label{eq:E^andB^}
\end{equation}
where $ C_1 $ and $ C_2 $ are constants of integration. Note that $ E_{\hat
r} $ and $
B_{\hat{~}} $ are related with $ E_{r} \equiv F_{tr} $ and $ B \equiv F_{r
\phi} $ through the
formulae:
\begin{equation}
E_{\hat r} = LN^{-1}( E_{r} + N^{\phi}B),
\hspace{5mm}
B_{\hat{~}} = K^{-1}LB.
\label{eq:EBandE^B^}
\end{equation}
Because $ g_{\phi \phi} (= K^2) $ should approach the radial coordinate
squared $ r^2 $
as $ r \rightarrow \infty $, the constant $ C_1 $ is identified with
the electric
charge $ Q_e $; whereas $ C_2 $ will be determined by the boundary condition $
N^{\phi}(\infty) = 0 $.
These will be explicitly shown in the following.

The nonvanishing components of the Einstein equation (\ref{eq:Ein}) take
the following
forms:
\begin{eqnarray}
R_{\hat{t}\hat{t}}
&=& LL'N^{-1}N'+L^2[(N^{-1}N')'+N^{-1}N'(N^{-1}N'+K^{-1}K')]-2\beta^2
\nonumber \\
&=& -2\Lambda + 8{\pi}G B_{\hat{~}}^2,
\label{eq:Rtt}\\
R_{\hat{\phi}\hat{\phi}}
&=&-LL'K^{-1}K'-L^2[(K^{-1}K')'+K^{-1}K'(N^{-1}N'+K^{-1}K')]-2\beta^2
\nonumber \\
&=& 2\Lambda + 8{\pi}G E_{\hat r}^2, \label{eq:Rpp}\\ R_{\hat{t}\hat{\phi}}
&=&L(\beta'+2K^{-1}K'\beta) = - 8{\pi}G E_{\hat r}B_{\hat{~}}, \label{eq:Rtp}\\
G_{\hat{r}\hat{r}}
&=& L^2N^{-1}N'K^{-1}K'+\beta^2
= -\Lambda + 4{\pi}G(B_{\hat{~}}^2-E_{\hat r}^2), \label{eq:Grr}
\end{eqnarray}
with $ \beta \equiv - \frac{1}{2}KLN^{-1}{N^{\phi}}' $.

Now we assume the SD (ASD) equation (\ref{eq:SD}) on the electric and
magnetic fields.
Substituting $ E_{\hat r} $ and
$ B_{\hat{~}} $ of (\ref{eq:E^andB^}) into the SD (ASD) equation
(\ref{eq:SD}) we have
\begin{equation}
N= \varepsilon K \tilde{N}^{\phi}, \hspace{5mm} \tilde{N}^{\phi} \equiv
N^{\phi}+C_2/C_1.
\label{eq:N=KN}
\end{equation}
{}From (\ref{eq:Grr}) and (\ref{eq:N=KN}) we can deduce the following equation:
\begin{equation}
L(N^{-1}N'+K^{-1}K')= \pm 2|\Lambda|^{1/2}. \label{eq:1stode1}
\end{equation}
In the following we will choose the positive sign without any loss of
generality. This leads
to the positive $ L $ at the infinity. It is easy to see that the remaining
equations
(\ref{eq:Rtt})-(\ref{eq:Rtp}) in the $ (t, \phi) $ space reduce to the
following pair of first-
order differential equations:
\begin{equation}
LN^{-1}N' = |\Lambda|^{1/2}-\varepsilon \beta, \hspace{5mm} LK^{-1}K' =
|\Lambda|^{1/2}+\varepsilon \beta, \label{eq:1stode2}
\end{equation}
and the one for $ \beta $:
\begin{equation}
(\beta+\varepsilon|\Lambda|^{1/2})(K^2\beta)'+ 8{\pi}GC_1^2K^{-1}K'=0.
\label{eq:1stode3}
\end{equation}
We can easily integrate (\ref{eq:1stode3}) to obtain $ \beta $ as a
function of $ K $:
\begin{equation}
\beta = \varepsilon|\Lambda|^{1/2} \bigl( \frac{\rho^2}{K^2} -1 \bigr),
\label{eq:beta}
\end{equation}
where $ \rho $ is implicitly defined as a function of $ K $ by the relation
\begin{equation}
\rho^2 + r_0^2 \ln|\frac{\rho^2- r_0^2}{r_0^2}|+q =K^2, \label{eq:rho}
\end{equation}
with $ q $ a constant of integration and $ r_0^2 \equiv
4{\pi}GC_1^2/|\Lambda| $. From
above equations we immediately obtain $ L $, $ N $ and $ \tilde{N}^{\phi} $:
\begin{equation}
L=|\Lambda|^{1/2}\frac{\rho^2-r_0^2}{\rho \rho'},
\hspace{1cm}
N=C|\Lambda|^{1/2}\frac{\rho^2- r_0^2}{K},
\hspace{1cm}
\tilde{N}^{\phi}=\varepsilon C|\Lambda|^{1/2}
\frac{\rho^2-r_0^2}{K^2},
\label{eq:L,N,N}
\end{equation}
with $ C $ a constant of integration.
We may set $ C =1 $ because the line element (\ref{eq:lineelement}) is
invariant under
the transformation $ t \rightarrow Ct $, $ N \rightarrow C^{-1}N $ and $
N^{\phi}
\rightarrow C^{-1}N^{\phi} $ and the constant $ C $ in $ N $ and $
\tilde{N}^{\phi} $ of
(\ref{eq:L,N,N}) can be absorbed into the time variable $ t $ and the
constant $ C_2/C_1
$ by suitable redefinitions of them.

We here impose a coordinate condition on our solution (\ref{eq:L,N,N}) to
see the
correspondence between our solution and the BTZ solution \cite{BTZ,BHTZ},
which will
clarify the physical properties of our solution.
The simplest choice
\footnote[1]
{Another choice, e.g., $ K = r $
may also be possible, but an invariant
volume element $ \omega^{\hat{t}} \wedge \omega^{\hat{r}} \wedge
\omega^{\hat{\phi}} =
KL^{-1}Ndtdrd\phi = \rho\rho'dtdrd\phi $ takes the simplest form $
rdtdrd\phi $ in our
choice.} of it will be attained by
putting $ \rho = r $ in (\ref{eq:rho}), i.e.,
\begin{equation}
K^2= r^2 + r_0^2 \ln|\frac{r^2- r_0^2}{r_0^2}|+q. \label{eq:K(r)}
\end{equation}
Note that $ K $ approaches $ r $ as $ r \rightarrow \infty $ and this
asymptotic behavior of
$ K $ allows one to identify $ C_1 $ with the electric charge $ Q_e $: $
C_1 \equiv Q_e $.
In the following we will simply put $ q = 0 $ in (\ref{eq:K(r)}) because it
gives
$ K^2 \rightarrow r^2 $ in the neutral limit $ Q_{e} \rightarrow 0 $. Note
that the
logarithmic function in
the right-hand side of (\ref{eq:K(r)}), which is proportional to $ Q_e^2 $,
stems from the
spacetime dimensionality. This is one of the peculiarities of the $ 2+1 $
dimensional
spacetime.

Since $ {\tilde N}^{\phi} $ of (\ref{eq:L,N,N}) approaches the constant
value $ \varepsilon
|\Lambda|^{1/2} $ as $ r \rightarrow \infty $ we see from (\ref{eq:N=KN})
and the boundary
condition $ N^{\phi}(\infty)= 0 $ that $ C_2 $ is given by $ C_2= \varepsilon
|\Lambda|^{1/2}Q_e $. Then we have
\begin{equation}
N^{\phi}= -\varepsilon \frac{4{\pi}GQ_{e}^2}{|\Lambda|^{1/2}}
\frac{ 1+\ln|\frac{r^2-r_0^2}{r_0^2}|}{K^2}. \label{eq:N^phi}
\end{equation}
Comparing the asymptotic form of $ N^{\phi} $ of (\ref{eq:N^phi}) with the
formula $
N^{\phi}= -J/(2r^2) $ of the BTZ solution \cite{BTZ,BHTZ}
we see that our solution has the following angular momentum $ J $:
\begin{equation}
J= \varepsilon \frac{8{\pi}GQ_{e}^2}{|\Lambda|^{1/2}}. \label{eq:J}
\end{equation}
The origin of this angular momentum can be considered to be the ``Poynting
pseudovector'' $ E_{\hat r}B_{\hat{~}} $, which gives the black hole the
nonzero angular
momentum $ J $. Our line element is given by
\begin{eqnarray}
ds^2 &=& -\frac{r^2}{K^2}
\bigl( -8{\pi}GQ_{e}^2 + |\Lambda|r^2 +\frac{J^2}{4r^2} \bigr) dt^2 +
\frac{dr^2}{-8{\pi}GQ_{e}^2 + |\Lambda|r^2 +\frac{J^2}{4r^2}} \nonumber \\ & &
\hspace{1cm} +K^2
\bigl[ -\frac{J}{2K^2} \bigl( 1+\frac{K^2-r^2}{r_0^2} \bigr) dt+d\phi \bigr]
^2.
\label{eq:ds^2 of sol}
\end{eqnarray}
This line element approaches that of the BTZ solution \cite{BTZ,BHTZ}
as $ r \rightarrow \infty $ since $ K $ approaches $ r $ in this limit, and
we see that our
solution has a mass $ M=8{\pi}GQ_{e}^2 $.

Since $ N^2 $ and $ L^2 $ vanish only at $ r = r_0 $, our solution is an
extreme black
hole with a horizon at $ r = r_0 $. The mass $ M $ and the angular momentum
$ J $ of
our solution obey a linear relation
$ M =|\Lambda|^{1/2}|J| $, which implies a saturation of the inequality $ M
\geq|\Lambda|^{1/2}|J| $ of BTZ \cite{BTZ,BHTZ}. In the neutral limit $ Q_e
\rightarrow 0 $
we have $ M \rightarrow 0 $ and $ J \rightarrow 0 $.
Furthermore we note that the $ tt $ component of the metric tensor
\begin{equation}
g_{tt} = -|\Lambda| \bigl( r^2 -2r_0^2- r_0^2
\ln|\frac{r^2-r_0^2}{r_0^2}| \bigr)
\label{eq:gtt)}
\end{equation}
vanishes only at $ r = \sqrt{2}r_{0} $ and $ r = \sqrt{2r_0^2-r_c^2} $,
where $ r_c $ is
defined by the equation:
\begin{equation}
K^2(r_c) = r_c^2 + r_0^2 \ln|\frac{r_c^2- r_0^2}{r_0^2}|=0. \label{eq:r_c)}
\end{equation}
We can show that $ r_c $ is in the region $ r_0 < r_c < \sqrt{2}r_0 $;
numerically we have
$ r_c/r_0 = 1.13069\cdots $. We call the region between $ r_0 $ and the
outer surface of
the infinite redshift $ r_{erg} \equiv \sqrt{2}r_0 $ the ergosphere \cite{HE}.

The electric field $ E_{\hat r} = Q_e/K $ and the magnetic field $
B_{\hat{~}} = \varepsilon
Q_e/K $ diverge at $ r =r_c $ because $ K $ vanishes at $ r =r_c $. Further
$ K $ is pure
imaginary in the region $ 0 < r < r_c $, which, however, simply shows that the
orthonormal 1-form basis
(\ref{eq:obasis}) should be replaced by more suitable one in this region.
In fact we have the following field strength 2-form:
\begin{equation}
F =-Q_e\frac{r}{r^2-r_0^2} dr \wedge
(dt-\varepsilon |\Lambda|^{-1/2}d\phi),
\label{eq:F}
\end{equation}
which has a singularity at $ r =r_0 $ but not at $ r =r_c $. The gauge
potential 1-form is
given by
\footnote[1]
{The temporal component of $ A $ in (\ref{eq:A}) has an asymptotic form $ -Q_e
\ln(r/r_0)dt $ as $ r \rightarrow \infty $. This expression agrees with the
gauge potential 1-
form proposed by BTZ \cite{BTZ,BHTZ}, though it is not a solution when $
JQ_e \neq 0 $.}
\begin{equation}
A =-\frac{Q_e}{2}\ln|\frac{r^2-r_0^2}{r_0^2}|
(dt-\varepsilon |\Lambda|^{-1/2}d\phi).
\label{eq:A}
\end{equation}
At $ r = 0 $ the field strength 2-form $ F $ and the gauge potential 1-form
$ A $ vanish
and we have no ``Dirac string'' because $ A_{\phi}(0) = 0 $.

{} From the formula for the scalar curvature $ R = 6\Lambda-16{\pi}GT $, we
see that our
spacetime has a constant negative curvature $ R = 6\Lambda $, which
coincides with
that of BTZ \cite{BTZ,BHTZ} with no Maxwell field.
We obtained this result from the tracelessness of our $ T^{\mu}_{~\nu} $,
i.e., $ T $ is
expressed as $ T \equiv T^{\mu}_{~\mu} = \frac{1}{2}
(B_{\hat{~}}^2 - E_{\hat r}^2) $ and this vanishes under the SD (ASD) equation
(\ref{eq:SD}).

The fact that both of the scalar curvature $ R $ and the invariant volume
element
$ \omega^{\hat{t}} \wedge \omega^{\hat{r}} \wedge \omega^{\hat{\phi}} =
rdtdrd\phi $
have no singularity at any value of $ r $ suggests that the singularities
at $ r = r_0 $ and $
r = r_c $ of the metric are coordinate singularities and they may be
removed by a suitable
redefinition of the coordinate system.
In fact we can see from (\ref{eq:gtt)}) that $ g_{tt} $ has singularity
only at $ r = r_0 $ and
further the singularities of the metric tensors $ g_{\mu\nu} $ and $
g^{\mu\nu} $, and
therefore the Riemann tensor, exist only at $ r = 0 $ or $ r = r_0 $. ( The
singularity at $ r =
0 $ appears only in $ g^{rr} $.) Therefore the singularity of the line
element (\ref{eq:ds^2
of sol}) at $ r = r_c $ is an apparent one. But we need a more careful
consideration about
the removability of singularities at $ r = r_0 $.

Although our scalar curvature $ R = 6\Lambda $ is the same as that of BTZ
\cite{BTZ,BHTZ} with no Maxwell field, the nonvanishing field strength
2-form (\ref{eq:F})
dwells in our spacetime.
Since we have
\begin{equation}
R^{\mu\nu}_{~~\rho\sigma}=\Lambda
(\delta^{\mu}_{\rho} \delta^{\nu}_{\sigma}
-\delta^{\mu}_{\sigma}\delta^{\nu}_{\rho})
+8{\pi}G \varepsilon^{\mu\nu\lambda} \varepsilon_{\rho\sigma\tau}
T^{\tau}_{{~}\lambda},
\label{eq:R=dd-dd+T}
\end{equation}
where $ \varepsilon $'s are the completely antisymmetric symbols and the
second term
in the right-hand side of (\ref{eq:R=dd-dd+T}) cannot be cast into the form
of the first one,
we see that our spacetime is not a symmetric one and it will not be
identified with an anti-
de Sitter space
in a four-dimensional spacetime by any embedding \cite{BTZ,BHTZ}.

Our solution describes the electrically charged extreme black hole with
mass $ M =
8{\pi}GQ_{e}^2 $ and angular momentum $ J = \pm
8{\pi}GQ_{e}^2/|\Lambda|^{1/2} $. It
has the event horizon at $ r = r_{0} $ and the ergosphere in the region $
r_{0} < r <
\sqrt{2}r_0 $. This extreme solution is a measure-zero solution in the $
(M, J, Q_e) $
space,
since these parameters are restricted in a curve in the parameter space.
We hope that our paper paves the way for the derivation of a generic
solution such as the
Kerr-Newman solution \cite{KN} in the four-dimensional spacetime.

\bfl {\bf Acknowledgments} \efl

One of the authors (M.K.) would like to thank Doctor R. Stark for helpful
discussions. He would also like to thank Professor Abdus Salam, the
International Atomic Energy Agency and UNESCO for hospitality at the
International Centre for Theoretical Physics, Trieste. The authors also
thank Professor K.S. Narain for reading the manuscript.

\end{document}